# On the beams of wavy metric tensors

A. LOINGER

Dipartimento di Fisica, Università di Milano

Via Celoria, 16 –  20133 Milano, Italy

**Summary.** – If we move at the same speed of a spatially limited train of an undulating metric tensor, a famous Serini's theorem will assure us that this train represents actually a flat spacetime region.

**1**. – The treatise of 1960 by Infeld and Plebanski contains an unsurpassed application of the Einstein-Infeld-Hoffmann method to the problem of gravitational radiation [1]. These authors prove that at each stage of approximation it is possible to choose a reference system for which *no* emission of gravity waves does happen. The present writer has recently published a simple, non-perturbative, general demonstration that *no* motion of point masses gives origin to emission of gravitational radiation [2].

I wish now to set forth another argument against the *physical* existence of gravitational waves, which utilizes suitably a remarkable Serini's theorem of 1919, generalized by Einstein and Pauli in 1943, and by Lichnerowicz in 1946 [3]. A perspicuous formulation of Serini's result has been given independently by Fock in his book on relativity theory [4]. In what follows I shall rest substantially upon Fock's ideas.

**2**. – For our purpose it is convenient to start from a Gaussian ("synchronous", according to Landau's terminology) expression of the spacetime interval [5]:

(2.1) $$\mathrm{d}s^2 = c^2\mathrm{d}t^2 - h_{\alpha\beta}\,(x,t)\mathrm{d}x^\alpha \mathrm{d}x^\beta, \quad (\alpha,\,\beta =1,2,3)\ ;$$

let us put

(2.2) $$\kappa_{\alpha\beta} := \frac{\partial h_{\alpha\beta}}{c\partial t}\ ;$$

$h_{\alpha\beta}$ and $\kappa_{\alpha\beta}$ are three-dimensional tensors. If $g_{jk}$, $(j,k=0,1,2,3)$, is the four-dimensional metric tensor and $g := \det\left\| g_{jk} \right\|$, we have

(2.3) $$\kappa^\alpha_\alpha := h^{\alpha\beta}\,\frac{\partial h_{\alpha\beta}}{c\partial t} = \frac{\partial}{c\partial t}\ln(-g)\ .$$

If $R_{jk}$ is the Ricci-Einstein tensor, the field equations for a mass tensor $T_{jk}$ which is rigorously equal to zero (with no singularities) can be written as follows:

(2.4) $$\equiv -\frac{1}{2c}\frac{\kappa}{\partial}\ -\frac{}{4}\kappa\ \kappa^\alpha\ \ 0\ ,$$

$$R_{0\alpha} \equiv \frac{1}{2}(\kappa^\beta_{\alpha;\beta} - \kappa^\beta_{\beta;\alpha}) = 0\ ,$$

(2.4") $$R_{\alpha\beta} \equiv \frac{1}{2c}\frac{\partial\kappa_{\alpha\beta}}{\partial t} + \frac{1}{4}(\kappa_{\alpha\beta}\,\kappa^\gamma_\gamma - 2\kappa^\gamma_\alpha\,\kappa_{\beta\gamma}) + \mathrm{P}_{\alpha\beta} = 0\ ,$$

where $\mathrm{P}_{\alpha\beta}$ is the three-dimensional analogue of $R_{jk}$, and we have applied tensor analysis to the three-dimensional space whose metric tensor is $h_{\alpha\beta}$.

Now, Serini's theorem concerns the interesting instances in which $g_{jk}$ is *time-independent* and regular. Then, in lieu of (2.4), (2.4'), (2.4") we have simply

(2.4 bis) $$R_{00} \equiv 0\ ,$$

(2.4'bis) $$R_{0\alpha} \equiv 0\ ,$$

(2.4"bis) $$R_{\alpha\beta} \equiv \mathrm{P}_{\alpha\beta} = 0\ ,$$

A basic result of tensor analysis tells us that if $P_{\alpha\beta}$ is equal to zero, also the three-dimensional curvature tensor $P_{\alpha\beta\gamma\delta}$ is equal to zero.

*Conclusion*: If the time-independent $g_{jk}$, $(j,k = 0,1,2,3)$, is regular everywhere and Minkowskian at the spatial infinity, the spatio-temporal interval d$s$ coincides with Minkowski's d$s$ . *Q.e.d.*

**3**. − Let us now consider a rather particular, but significant, example: a finite beam − i.e. a spatially limited train − of hypothetical gravity waves, which arrives from a remote spatial region. Remembering that in *general* relativity there is *no* limitation for the values of the velocities of the reference frames, we can suppose to follow our beam in its motion by travelling at its same speed. But then the beam will be essentially seen by us as a spatially oscillatory gravity field *at rest*. Accordingly, its metric tensor $g_{jk}$ will be practically *time-independent* − in our reference system. By applying Serini's theorem, we conclude that our finite undulatory train represents actually a flat Minkowskian spacetime region, i.e. a region with four-dimensional curvature tensor equal to zero. *Our gravity radiation has revealed itself to be a phantom, i.e. a mere coordinate undulation.*

APPENDIX

**On a wrong conviction**

Many physicists think that without gravitational waves, one would have to explain an instantaneous propagation of a change in the metric over the whole universe simply obtained by changing the distribution of mass or stress in a physical system. This conviction is erroneous. In reality, the physical non-existence of the gravity waves is *fully consistent* with the fundamental principles of relativity theory: the Einstein field equations are time-symmetrical, and therefore the discarding of time-unsymmetrical solutions is quite legitimate. Analogously, Maxwell equations are time-symmetrical: the physical existence of electromagnetic waves is only a theoretically valid possibility, not a theoretical necessity: the real existence of the e.m. waves is an *experimental* fact. Any accelerated charge gives out e.m. waves, whose existence can be ascertained in any reference frame.

In Einstein theory and in Maxwell theory sources and fields are inseparable from each other, and it is impossible to distinguish cause from effect; but in Einstein theory the *physical* existence of the gravity waves is *not* a theoretically valid possibility.